\documentstyle[prl,aps,epsfig,multicol]{revtex}
\begin{document}

\title{Exchange Splitting and Charge Carrier Spin Polarization in EuO}
\author{P.~G.~Steeneken$^1$, L.~H.~Tjeng$^{1,2}$, I.~Elfimov$^1$,
G.~A.~Sawatzky$^1$, G.~Ghiringhelli$^3$, N.~B.~Brookes$^4$, and
D.-J.~Huang$^5$} 
\address{$^1$ Solid State Physics Laboratory, Materials
Science Centre, University of Groningen, Nijenborgh 4, 9747~AG Groningen,
the Netherlands.
\\$^2$ II. Physikalisches Institut, Universit\"at zu K\"oln,
Z\"ulpicher Stra\ss e 77, D-50937 K\"oln, Germany.
\\$^3$ Dipartimento di Fisica, INFM - Politecnico di
Milano, piazza Leonardo da Vinci 32 , 20133 Milano, Italy. 
\\$^4$ European
Synchrotron Radiation Facility, B.P. 220, 38043 Grenoble Cedex, France.
\\$^5$ Synchrotron Radiation Research Center, No. 1 R\&D VI, Hsinchu
Science-based Industrial Park, Hsinchu 30077, Taiwan.} 
\date{\today}
\maketitle
\begin{abstract}
High quality thin films of the ferromagnetic semiconductor EuO have been
prepared and were studied using a new form of spin-resolved spectroscopy. We
observed large changes in the electronic structure across the Curie and
metal-insulator transition temperature. We found that these are caused by
the exchange splitting of the conduction band in the ferromagnetic state,
which is as large as 0.6~eV. We also present strong evidence that the
bottom of the conduction band consists mainly of majority spins. This implies
that doped charge carriers in EuO are practically fully spin polarized.
\end{abstract}
\begin{multicols}{2}
EuO is a semiconductor with a band gap of about 1.2~eV and is one of the
very rare ferromagnetic oxides \cite{review1,review2}. Its Curie temperature
($T_c$) is around 69 K and the crystal structure is rocksalt (fcc) with a
lattice constant of 5.144 \AA. Eu-rich EuO becomes metallic below $T_c$ and
the metal-insulator transition (MIT) is spectacular: the resistivity drops
by as much as 8 orders of magnitude \cite{Oliver,Shapira}. Moreover, an
applied magnetic field shifts the MIT temperature considerably, resulting in
a colossal magnetoresistance (CMR) with changes in resistivity of up to 6
orders of magnitude \cite{Shapira}. This CMR behavior in EuO is in fact more
extreme than in the now much investigated La$_{1-x}$Sr$_x$MnO$_3$ materials
\cite{Ramirez,Imada}. Much what is known about the basic electronic
structure of EuO dates back to about 30 years ago and is based mainly on
optical measurements \cite{Busch,Freiser,Schoenes} and band structure
calculations \cite{Cho}. With the properties being so spectacular, it is
surprising that very little has been done so far to determine the electronic
structure of EuO using more modern and direct methods like electron
spectroscopies.

Here we introduce spin-resolved x-ray absorption spectroscopy, a new type of
spin-resolved electron spectroscopy technique to study directly the
conduction band of EuO where most of the effects related to the MIT and CMR
are expected to show up. Spin-resolved measurements of the conduction band
could previously only be obtained by spin-polarized inverse photoemission
spectroscopy. Spin-resolved x-ray absorption spectroscopy is an alternative
technique which is especially well suited to study ferromagnetic oxides, a
currently interesting broad class of materials. Using this technique we
observed large changes in the conduction band across $T_c$ and we were able
to show experimentally that these are caused by an exchange splitting of the
conduction band below $T_c$. Moreover, we found that this splitting is as
large as 0.6 eV and show that the states close to the bottom of the
conduction band are almost fully spin-polarized, which is very interesting
for basic research in the field of spintronics.

The experiments were performed using the helical undulator \cite{Elleaume94}
based beamline ID12B \cite{Goulon95} at the European Synchrotron Radiation
Facility (ESRF) in Grenoble. Photoemission and Auger spectra were recorded
using a 140~mm mean radius hemispherical analyzer coupled to a mini-Mott
25~kV spin polarimeter \cite{Ghiringhelli99}. The spin detector had an
efficiency (Sherman function) of 17\%, and the energy resolution of the
electron analyzer was 0.7~eV. The photon energy resolution was set at
0.2~eV. The measurements were carried out at normal emission with respect to
the sample surface and at an angle of incidence of the x-rays of 60$^\circ$.
The sample was magnetized remanently in-plane using a pulsed magnetic coil,
the magnetization direction was alternated to eliminate the effect of
instrumental asymmetries \cite{Getzlaff98}. The pressure of the spectrometer
chamber was better than 1x10$^{-10}$~mbar.

EuO samples with a film thickness of $\approx$ 200 \AA\ were grown {\it
in-situ} by evaporating Eu metal from a Knudsen cell at a rate of $\approx$
3 \AA\ per minute in an oxygen atmosphere of 1x10$^{-8}$~mbar on top of a Cr
covered, chemically polished Al$_2$O$_3$ substrate kept at 280$^{\circ}$C.
In a separate experiment in the Groningen laboratory, we have verified that
this recipe provides us with high quality polycrystalline EuO films. Valence
band and core level x-ray photoemission spectroscopy (XPS) show no
detectable presence of Eu$^{3+}$ ions, and ultra-violet photoemission
experiments reveal that there is no detectable density of states at the
Fermi level, demonstrating that we have indeed obtained nearly
stoichiometric semiconducting EuO without detectable traces of Eu$_2$O$_3$,
Eu$_3$O$_4$, or Eu metal. The left panel of Fig.1 displays the
magneto-optical Kerr-rotation as a function of temperature on our films using 
a He-Ne laser ($h\nu$ = 1.96~eV) and shows that these films have indeed the 
correct T$_c$ of 69 K. The right panel of Fig.1 depicts the resistivity of the 
films as a function of temperature, and it demonstrates clearly the presence of a
metal-insulator transition at T$_c$. The very large change 
\begin{figure}  
\narrowtext
\centerline{\epsfig{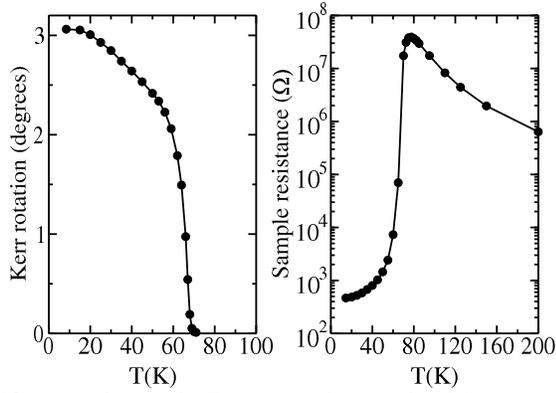}} \caption{Left panel: Remanent
longitudinal Kerr-rotation of a 50 nm EuO thin film as a function of
temperature using p-polarized light at $h\nu$ = 1.96~eV and
$\theta_{in}$ = 45$^\circ$. Right panel: Metal-insulator transition in
the temperature dependent resistivity of a EuO film.}
\end{figure}
\vspace{-0.3cm}
\noindent in resistivity, namely 5 orders in magnitude, indicates that the
carrier concentration due to oxygen defects, i.e. the off-stoichiometry, is
of the order of 0.3\% or less \cite{Shapira}.To study the conduction band of
EuO, we use O $K$-edge x-ray absorption spectroscopy (XAS). This technique
probes the O $2p$ character of the conduction band, which is present because
of the covalent mixing between the O $2p$ and Eu $5d$-$6s$ orbitals. Fig.2
displays the O $K$ XAS spectra, recorded by collecting the total electron
yield (sample current) as a function of photon energy. Large changes over a
wide energy range can be clearly seen between the spectra taken above and
below T$_c$. The low temperature spectrum contains more structures and is
generally also broader. We note that the spectra also show a very small
feature at 529.7~eV photon energy, with a spectral weight of not more than
0.1\% relative to the entire spectrum. Since the intensity of this feature
is extremely sensitive to additional treatments of
the sample surface it is probably related to surface states\cite{Schiller1}
or imperfections at the surface. 

It is evident that the spectral changes across T$_c$ cannot be
explained by a phonon mechanism, since the changes  
\begin{figure} 
\narrowtext 
\centerline{\epsfig{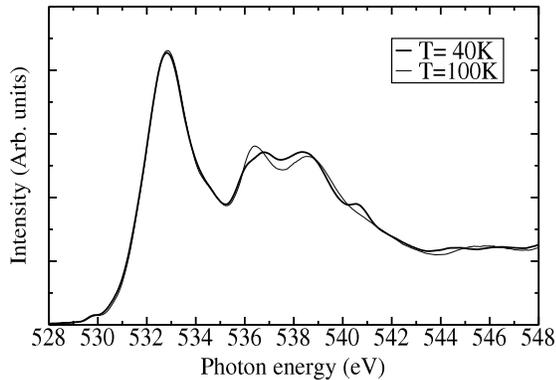}} 
\caption{O $K$ x-ray absorption spectrum of EuO, above (thin solid line) and 
below (thick solid line) the Curie temperature ($T_c$ = 69 K).} 
\end{figure} 
\noindent involve more than a
simple broadening, and above all, since the spectrum becomes broader
upon temperature lowering. Because there are also no changes in the
crystal structure across T$_c$, we attribute the spectral changes to
the appearance of a spin-splitting in the Eu $5d$ like conduction band
below T$_c$. To prove this, we have to determine the spin-polarized
unoccupied density of states. To this end, we measured the O $K$ XAS
spectrum no longer in the total electron yield mode, but in a partial
electron yield mode in which we monitor the O $KL_{23}L_{23}$ Auger peaks
that emerge at a constant kinetic energy from the XAS process. By
measuring the spin-polarization of this Auger signal while scanning
the photon energy across the O $K$ edge, we can obtain the {\it
spin-resolved} O $K$ XAS spectrum.

The underlying concept of this new type of experiment is illustrated
in Fig.3. This figure shows the O $1s$ core level, the occupied O $2p$
valence band and the unoccupied Eu $5d$-$6s$ conduction band.
Quotation marks indicate that due to covalent mixing the conduction
band also has some O $2p$ character and the valence band some Eu
$5d$-$6s$ character. This mixing allows an x-ray to excite an O $1s$
electron to the conduction band, leaving a spin-polarized core hole if
the conduction band is spin-polarized (middle panel Fig.3). The
subsequent $KL_{23}L_{23}$ Auger decay of the XAS state leads to
O($2p^4$) like final states \cite{Tjeng}, and the outgoing Auger
electron will now also be spin-polarized (right panel Fig.3). Unique
to a $KL_{23}L_{23}$ Auger decay is that the entire two-hole final
states are of pure singlet ($^1S$ and $^1D$) symmetry since the
triplet $^3P$ transitions are forbidden by Auger selection rules
\cite{Saw,Fug}. This implies that the O $KL_{23}L_{23}$ Auger electrons 
will have a degree of spin-polarization which is equal to that
 of the conduction band, but which has an
opposite sign due to the singlet character of the Auger transition. Thus,
the measurement of the spin of the O $KL_{23}L_{23}$ Auger electrons across
the O $K$ edge will reflect the spin-polarization of the unoccupied conduction
band states. We note that spin-resolved XAS is different from 
\begin{figure}  
\narrowtext
\centerline{\epsfig{file=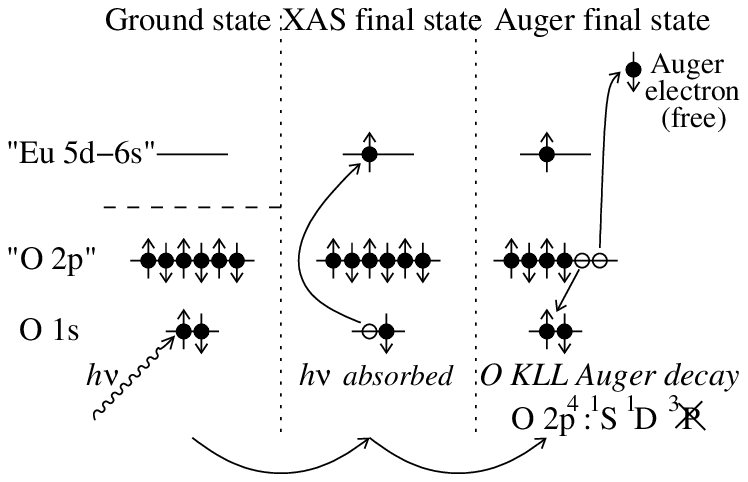}} \caption{Mechanism of
spin-resolved x-ray absorption spectroscopy. The spin of the outgoing
O $KL_{23}L_{23}$ Auger electron is opposite to the spin of the
electron that is excited in the x-ray absorption process. This scheme
illustrates the observation of a spin-up conduction band state.}
\end{figure}
\noindent a
magnetic circular dichroism (MCD) experiment. In the latter the helicity of
the circularly-polarized light is varied and the dichroic signal contains a
more convoluted information about the spin and the orbital moments of the
unoccupied states \cite{Tanaka}. We also note that the availability of EuO
in thin film form is crucial for the measurement of spin-polarized electron
spectroscopies because the remanent magnetic field created by the very small
amount of material involved is negligibly small, and thus a perturbation of
the trajectories of the emitted electrons can be avoided. As an
illustration, we show in Fig.4 a small selection of the photoelectron
spectra which we have taken from the EuO valence band and the O
$KL_{23}L_{23}$ Auger as a function of photon energy at T=20 K. The left
panel displays the unpolarized spectra while the right panel gives the
difference spectra between the spin-up and spin-down channels (the spin-up
direction is parallel to the magnetization direction). We can clearly
distinguish the narrow Eu $4f^7$$\rightarrow$$4f^6$ photoemission (PES) peak
in the valence band spectrum \cite{Eastman,Sattler}, and observe that its
spin-polarization is about 50\%, indicating that the remanent magnetization
of the EuO films at this temperature is only half of the saturation
magnetization. This remanence is confirmed by analyzing the magnetic
circular dichroism measurements at the Eu $M_{45}$ ($3d$$\rightarrow$$4f$)
photoabsorption edges and is comparable to magnetization measurements on EuO
films \cite{Iwata}.

It is interesting to see the strong photon energy dependence of the
magnitude and spin-difference of the O $KL_{23}L_{23}$ Auger. By measuring
these spin-resolved O $KL_{23}L_{23}$ Auger spectra across the entire O $K$
edge region with closely spaced photon energy intervals, we can construct an
accurate spin-resolved O $K$ XAS spectrum of EuO. The results are shown in
the top panel of Fig.5. Here a normalization for full sample magnetization
has been made using the measured spin-polarization of the 
\begin{figure}  
\centerline{\epsfig{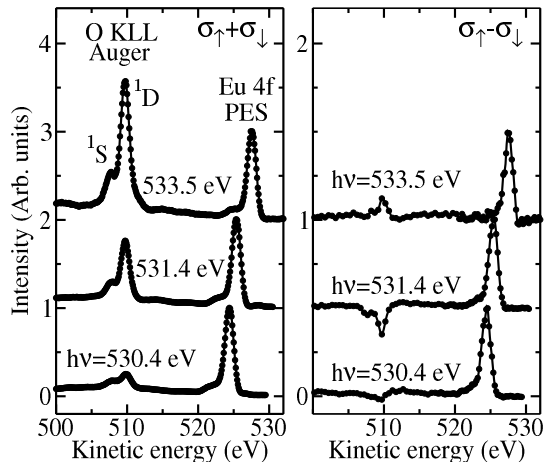}} \caption{Left panel:
Spin-integrated valence band photoemission and O $KL_{23}L_{23}$
Auger spectra of EuO. Right panel: Difference spectra between the
spin-up and spin-down channels for the valence band and O
$KL_{23}L_{23}$ Auger.}
\end{figure}
\noindent Eu $4f$. We
observe an almost rigid splitting between the spin-up and spin-down peaks
near the bottom of the conduction band, which is as large as 0.6 eV.
Extrapolation of the data strongly suggests a very high spin-polarization at
the bottom of the band. It is also interesting to note that the spin
polarized features at 536.0 eV and 536.8 eV correspond with features in the
low temperature XAS scan of Fig.2. and that above T$_c$ these peaks seem to
merge into one feature at 536.4 eV, indicating that the changes in the
density of states below T$_c$ can indeed largely be attributed to a
spin-splitting, which is a shift of the spin-up band to lower energy and the
spin-down band to higher energy. However at higher energies the
spin-behavior seems to be less simple: the spin-up feature at 540.7 eV, for
example, does not seem to have a spin-down counterpart, possibly due to the
near presence of the $4f^7$$\rightarrow$$4f^8$ electron addition peak in the
conduction band. To strengthen our understanding of the experimental
findings, we have also carried out band structure calculations in the local
spin density approximation including the on site Hubbard U (LSDA+U)
\cite{Anisimov} for EuO in the ferromagnetic state. The bottom panel 
\begin{figure}  
\centerline{\epsfig{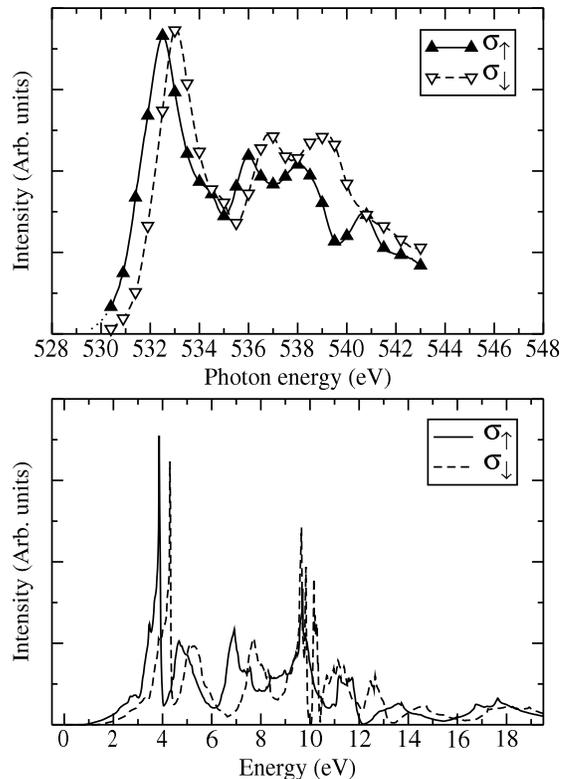}} 
\caption{Top panel: spin-resolved O $K$ x-ray absorption spectrum of EuO
taken at 20K. Bottom panel: spin-resolved unoccupied O 2p partial density of
states from LSDA+U calculations (U = 7.0 eV). The zero of energy corresponds
to the top of the valence band. For comparison: the difference between the O
$1s$ XPS onset and the valence band onset is 528.5$\pm$0.5 eV. The O $1s$
XPS onset is taken as being 1 eV below the peak value. }
\end{figure}
\noindent of
Fig.5 displays the results for the spin-resolved unoccupied O $2p$ partial
density of states. The agreement between this mean-field theory and the
experiment is remarkable: the approximate position and spin-splitting of
most peaks is well reproduced, including the more intricate features that
arise at about 10 eV above the Fermi level which are associated with the
unoccupied 4$f$ states. Recently we have become aware of calculations that
obtain results similar to ours\cite{Schiller2,Schiller3}. When comparing XAS
with band structure calculations it should be noted that the interaction
between the core hole and the conduction band electron can in principle lead
to exitonic effects in the XAS spectra. However these effects generally lead
to sharp white lines which we do not observe here. Moreover because the core
hole is on the oxygen atom while the conduction band consists of Eu
orbitals, the interaction with an O 1$s$ core hole is so small as compared
to the 5$d$ band width that its effect will be negligible. This conclusion
is also based on a vast amount of data on the O $K$ edge XAS of transition
metal oxides.
These experimental results clearly demonstrate that the large changes
observed in the conduction band structure below T$_c$ in Fig.2 are caused by
a spin-splitting. We attribute this splitting to the direct exchange
interaction between the localized $4f$ moments and the delocalized $5d$-$6s$
conduction band states. We note that these measurements suggest that the red
shift of the optical absorption edge is also due to this spin-splitting
rather than to a broadening of the conduction band. These results support
the following picture for the metal-insulator transition in Eu-rich
EuO\cite{Oliver}. Above $T_c$, defect or impurity states have their
energy levels located slightly below the bottom of the conduction band, and
the material behaves like a semiconductor: the resistivity decreases with
increasing temperatures as a result of a thermal activation of the electrons
from the defect states into the conduction band. Below $T_c$, the
conduction band splits due to the exchange interactions, and the defect
states now fall into the conduction band. The electrons of these defects
can then propagate in the spin-polarized bottom of the conduction band
without needing any activation energy, and the system behaves like a metal.
As we estimate the depolarization of the conduction band due to spin-orbit
coupling ($\xi_{5d}$ = 0.067~eV) to be small (less than 5\%), we expect the
doped charge carriers in ferromagnetic EuO to be almost fully spin
polarized, an observation that is very interesting for fundamental research
projects in the field of spintronics.
To conclude, our experiments have revealed large changes in the
conduction band states of EuO if the temperature is varied across
$T_c$. Using new spin-resolved measurements we have shown that these
changes are caused by a splitting between the spin-up and spin-down
unoccupied density of states. This exchange splitting is appreciable,
about 0.6~eV. From this we conclude that electron doped EuO in the
ferromagnetic state will have charge carriers with an almost 100\%
spin-polarization.

We would like to thank K. Larsson, J.~C. Kappenburg, R. Hesper, M.~V. Tiba and
R. Bhatia for their skillful assistance and D.~I. Khomskii, F.~M.~F. de Groot
and P.~D. Johnson for helpful discussions.  This work was supported by the
Netherlands Foundation for Fundamental Research on Matter (FOM) with financial
support from the Netherlands Organization for Scientific Research (NWO).

\narrowtext

\end{multicols}

\vspace{-0.5cm}
\begin{references}
\vspace{-1.5cm}
\bibitem{review1} for a review see:  "Electronic Conduction in Oxides" by
     N.~Tsuda, K.~Nasu, A.~Yanase, and K.~Siratori, Springer Series in
     Solid-State Sciences 94, (Springer Verlag, Berlin 1991).

\bibitem{review2} for a review see: A.~Mauger and C.~Godart,
     Phys. Rep. {\bf 141}, 51 (1986).

\bibitem{Oliver} M.~Oliver {\it et al.},
     Phys. Rev. B {\bf 5}, 1078 (1972).

\bibitem{Shapira} Y.~Shapira, S.~Foner, and T.B.~Reed,
     Phys. Rev. B {\bf 8}, 2299 (1973); {\it ibid.} 2316 (1973).

\bibitem{Ramirez} for a review see: A.P.~Ramirez,
     J. Phys.: Conden. Matter {\bf 9}, 8171 (1997).

\bibitem{Imada} for a review see: M.~Imada, A.~Fujimori, and Y.~Tokura,
     Rev. Mod. Phys. {\bf 70}, 1039 (1998).

\bibitem{Busch} G.~Busch, P.~Junod, and P.~Wachter,
     Phys. Lett. {\bf 12}, 11 (1964).

\bibitem{Freiser} M.J.~Freiser {\it et al.},
     Helv. Phys. Acta {\bf 41}, 832 (1968).

\bibitem{Schoenes} J.~Schoenes and P.~Wachter,
     Phys. Rev. B {\bf 9}, 3097 (1974).

\bibitem{Cho} S.J.~Cho, Phys. Rev. B {\bf 1}, 4589 (1970).

\bibitem{Elleaume94} P.~Elleaume,
    J. Synchrotron Radiat. {\bf 1}, 19 (1994).

\bibitem{Goulon95} J.~Goulon {\it et al.},
    Physica B {\bf 208}, 199 (1995).

\bibitem{Ghiringhelli99} G.~Ghiringhelli, K.~Larsson and N.B.~Brookes,
    Rev. Sci. Instrum. {\bf 70}, 4225 (1999).

\bibitem{Getzlaff98} Equation (15) in M.~Getzlaff {\it et al.},
    Rev. Sci. Instrum. {\bf 69}, 3913 (1998).

\bibitem{Schiller1} R.~Schiller and W.~Nolting,
    Phys. Rev. Lett. {\bf 86}, 3847 (2001).

\bibitem{Tjeng} L.H.~Tjeng {\it et al.}, in "Strong Correlation and
    Superconductivity", edited by H.~Fukuyama, S.~Maekawa, and
    A.~P.Malozemoff, Springer Series in Solid State Sciences,
    Vol. {\bf 89} (Springer Verlag, Berlin 1989), p. 33.

\bibitem{Saw} G.A.~Sawatzky, lecture notes on Auger Spectroscopy, 
   http://vsf1.phys.rug.nl/$\sim$surphys/lectures/auger.pdf

\bibitem{Fug} J.C.~Fuggle, in "Electron Spectroscopy: Theory, Techniques and
    Applications", edited by C.R.~Brundle, and A.D.~Baker
    (Academic Press, London and New York 1981), Vol. 4, p. 85.

\bibitem{Tanaka} A.~Tanaka, private communication.

\bibitem{Eastman} D.E.~Eastman, F.~Holtzberg, and S.~Methfessel,
     Phys. Rev. Lett. {\bf 23}, 226 (1969).

\bibitem{Sattler} K.~Sattler and H.C.~Siegmann,
     Phys. Rev. Lett. {\bf 29}, 1565 (1972);
     Z. Phys. B. {\bf 20}, 289 (1975).

\bibitem{Iwata} N.~Iwata {\it et al.},
     J. Phys. Soc. Jpn. {\bf 69}, 230 (2000).

\bibitem{Anisimov} V.I.~Anisimov, J.~Zaanen, and O.K.~Andersen,
     Phys. Rev. B {\bf 44}, 943 (1991).

\bibitem{Schiller2} R.~Schiller and W.~Nolting,
     Solid State Commun. {\bf 118}, 173 (2001).

\bibitem{Schiller3} R.~Schiller, Dissertation, Humboldt-Universit\"at zu
Berlin (2000) http://dochost.rz.hu-berlin.de/dissertationen/
schiller-roland-2000-11-01/PDF/Schiller.pdf

\end{references}
\end{document}